\documentclass[twocolumn,showpacs,aps]{revtex4}

\usepackage{epsfig}

\def\cal#1{{\cal #1}}

\def\m@th{\mathsurround=0pt}
\def\n@space{\nulldelimiterspace=0pt \m@th}
\def\biggg#1{{\mbox{$\left#1\vbox to 20.5pt{}\right.\n@space$}}}

\def\beginenum{\begin{enumerate}}
\def\endenum{\end{enumerate}}
\def\bitem{\begin{itemize}}
\def\eitem{\end{itemize}}
\def\bray{\begin{array}}
\def\eray{\end{array}}
\def\begindoc{\begin{document}}
\def\enddoc{\end{document}}
\def\bq{\begin{equation}}
\def\eq{\end{equation}}
\def\bqy{\begin{eqnarray}}
\def\eqy{\end{eqnarray}}
\def\bqyn{\begin{eqnarray*}}
\def\eqyn{\end{eqnarray*}}
\def\bc{\begin{center}}
\def\ec{\end{center}}
\def\bfll{\begin{flushleft}}
\def\efll{\end{flushleft}}
\def\bflr{\begin{flushright}}
\def\eflr{\end{flushright}}
\newcommand{\Avec}{\mbox{\boldmath $A$}}
\newcommand{\Bvec}{\mbox{\boldmath $B$}}
\newcommand{\Evec}{\mbox{\boldmath $E$}}
\newcommand{\Fvec}{\mbox{\boldmath $F$}}
\newcommand{\Gvec}{\mbox{\boldmath $G$}}
\newcommand{\Rvec}{\mbox{\boldmath $R$}}
\newcommand{\Uvec}{\mbox{\boldmath $U$}}
\newcommand{\Vvec}{\mbox{\boldmath $V$}}
\newcommand{\evec}{\mbox{\boldmath $e$}}
\newcommand{\jvec}{\mbox{\boldmath $j$}}
\newcommand{\kvec}{\mbox{\boldmath $k$}}
\newcommand{\nvec}{\mbox{\boldmath $n$}}
\newcommand{\uvec}{\mbox{\boldmath $u$}}
\newcommand{\vvec}{\mbox{\boldmath $v$}}
\newcommand{\wvec}{\mbox{\boldmath $w$}}
\newcommand{\xvec}{\mbox{\boldmath $x$}}
\newcommand{\omegavec}{\mbox{\boldmath $\omega$}}
\newcommand{\Omegavec}{\mbox{\boldmath $\Omega$}}

\usepackage{amsfonts}
\usepackage{amsmath}
\usepackage{amssymb}

\begin{document}

\title{Vortex Bubble Formation in Pair Plasmas}

\author{V.I. Berezhiani$^{\,1,2}$, N.L. Shatashvili$^{\,1,3}$,
S.M. Mahajan$^{\,4}$ and B.N. Aleksi\'c$^{\,5,6}$}

\affiliation{$^{1}$Andronikashvili Institute of Physics, Tbilisi
0177, Georgia }

\affiliation{$^{2}$School of Physics, Free University of Tbilisi,
Georgia}

\affiliation{$^{3}$Department of Physics, Faculty of Exact and
Natural Sciences, Ivane Javakhishvili Tbilisi State University,
Tbilisi 0179, Georgia}

\affiliation{$^{4}$Institute for Fusion Studies, The University of
Texas at Austin, Austin,Tx 78712}

\affiliation{$^{5}$Institute of Physics Belgrade, University of
Belgrade, Belgrade, Serbia}

\affiliation{$^{6}$ Texas A\&M Univeristy at Qatar, Doha, Qatar}

\begin{abstract}
It is shown that delocalized vortex solitons in relativistic pair
plasmas with small temperature asymmetries can be unstable for
intermediate intensities of the background electromagnetic field.
Instability leads to the generation of ever--expanding cavitating
bubbles in which the electromagnetic fields are zero. The
existence of such electromagnetic bubbles is demonstrated by
qualitative arguments based on a hydrodynamic analogy, and by
numerical solutions of the appropriate Nonlinear Schr\"odinger
equation with a saturating nonlinearity.

\end{abstract}


\maketitle

Nonlinear dynamics of vortex solitons, structures that permeate
different fields of physics, like Nonlinear Optics, Bose-Einstein
Condensate and Plasma Physics, has received considerable attention
in the recent past \cite{Kivshar}. \  The vortex solitons,
usually, emerge as solutions of a generalized (2+1)D-dimensional
nonlinear Schr\"odinger equation (NSE) with a local nonlinearity.
The NSE with a defocusing (for instance, cubic) nonlinearity
admits stable vortex soliton solutions. These solutions, with an
angular $2\pi$ phase ramp, and appearing as local dark minima in
an otherwise bright background, are viewed as the most fundamental
two-dimensional solitons embodied in the NSE. The NSE with a
focusing nonlinearity also supports localized vortex solitons with
phase dislocations surrounded by one or many bright rings. The
latter, however, are unstable against symmetry-breaking
perturbations that can cause the breakup of rings into filaments.

New and interesting behavior arises when the nonlinearity can
change its character (as a function of the amplitude, for
instance) from focusing to defocusing or vice versa. A
representative example is a sign -changing, cubic-quintic (CQ)
nonlinearity (focusing for lower and defocusing for higher
intensities). The NSE with a CQ nonlinearity can, under certain
conditions, support existence of stable localized as well as
delocalized vortex soliton solutions (see for details
\cite{Ber1}). Though CQ anzats is frequently invoked to model
light beam propagation in various optical media \cite{malomed}
(like non-Kerr crystals, chalcogenide glasses, semiconductors and
so on), its validity to describe  these materials does require
additional explorations.

In contrast to the optical media, an NSE with a
focusing-defocusing nonlineartity can naturally  emerge from the
basic physics models when one investigates the propagation of
large amplitude electromagnetic pulses  in a plasma
\cite{PlasmaSat}. \ In some of our recent publications
\cite{Nana},\cite{Ber2}, we showed that the dynamics of the short
intense electromagnetic pulses propagating in a relativistic pair
plasmas with small temperature asymmetry does, indeed, obey an NSE
with, what may be called, a saturation nonlinearity. Though always
positive (unlike the CQ nonlinearity), the saturation nonlinearity
vanishes for high intensities, and thus combines the
focusing-defocusing natures.

The particular form of the saturation function in the NSE derived
in \cite{Nana},\cite{Ber2}, leads to the existence of localized as
well as delocalized optical vortex soliton solution (we term these
solitons as LOVS and OVS respectively). The stability of LOVS
(both single and multi-charged) has been investigated in detail
and the stability areas were established. Although the stability
of the OVS has been established for high intensities of the
background fields \cite{Ber2}, for intermediate intensities the
vortex solitons can be unstable. Indeed Kim \textit{et} \textit{al
}\cite{Kim} (see also \cite{Berloff}) demonstrated numerically
that delocalized vortices described by NSE with CQ saturation
turns out to be unstable for certain intermediate intensities
leading to the formation of an ever-expanding circular ring. One
is naturally faced with the question if similar behavior would
pertain for the saturation nonlinearity. Unfortunately the
standard linear modal stability analysis is inadequate to deal
with non-exponential growth that could be associated with
non-Hermitian (non-self-adjoint) linear operators. To determine
the stability of the  OVS in a pair plasma, therefore, we resort
to a numerical solution of the NSE.

\bigskip

The nonlinear (2+1)D evolution of an electromagnetic (EM) pulse
propagating in an arbitrary pair plasma with temperature asymmetry
can be described by an NSE. The dimensionless evolution equation
for the slowly varying envelope of the vector potential (in
co-moving with group velocity frame) reads \cite{Ber2}:
\begin{equation}
i\frac{\partial A}{\partial t}+\nabla_{\perp}^{2}\,A+f(|A|^{2})A=0.
\label{B1}
\end{equation}
The nonlinearity function
\begin{equation}
f(|A|^{2})=\frac{|A|^{2}}{(1+|A|^{2})^{2}}\ ,
\label{B2}
\end{equation}
has a very unusual feature -- in the ultra-relativistic limit \
($|A|^{2} \gg1$) \ it tends to be $0$ and the nonlinear term vanishes.

\bigskip

Before proceeding further, we summarize the steps that led to
\eqref{B1}. We had started with a circularly polarized EM pulse \
$\sim \widehat
{A}\,(\hat{\mathbf{x}}+i\hat{\mathbf{y}})\,\mathrm{{exp}(ik_{0}z-\omega_{0}t)}$
with mean frequency $\omega_{0}$ and mean wave number $k_{0}$\ and
where $\widehat {A}$ \ is the slowly varying amplitude. In
Eq.(\ref{B1}), \ $\nabla_{\perp}^{2}=\partial ^{2}/\partial
x^{2}+\partial^{2}/\partial y^{2}$ \ is the diffraction operator,
and various dimensionless quantities have been defined as: \
$A=|e|\widehat{A}/(mG(T_{0}^{-})\,c^{2})$, \ $r=\left(
\omega_{e}/c\,\epsilon k^{1/2}\right)  r$, \
$t=(\omega_{e}^{2}/2\,\omega_{0}\,\epsilon^{2}k)t$; where
$\omega_{e}=(4\pi e^{2}n_{0}/m)^{1/2}$ is the electron Langmuir
frequency and $m$ is the electron mass. The charges $q^{\pm}$ and
masses $m^{\pm}$ of positive and negative ions are assumed to be
same (we mainly concentrate on the specific case of an
electron-positron plasma, i.e. $q^{+}=e^{+}=q^{-}=-e^{-}=|e|$ and
$m^{+}=m^{-}=m$). The equilibrium state of the system is
characterized by an overall charge neutrality $n_{0}^{+}
=n_{0}^{-}=n_{0}$ where $n_{0}^{+}$ and $n_{0}^{-}$ are the
unperturbed number densities of the positive and negative species
respectively. The background temperatures of plasma species are
$T_{0}^{\pm}$ ($T_{0}^{+}\neq T_{0}^{-}$) and
$m\,G(z^{\pm})=m\,K_{3}(z^{\pm})/K_{2}(z^{\pm})$ is the
\textquotedblright effective mass\textquotedblright,
[$z^{\pm}=mc^{2}/T^{\pm}$], where $K_{\nu}$ are the modified
Bessel functions. For the nonrelativistic temperatures
($T^{\pm}\ll mc^{2}$) $G^{\pm}=1+5T^{\pm}/2mc^{2}$ and for the
ultra-relativistic temperatures ($T^{\pm}\gg m_{\alpha}c^{2}$)
$G^{\pm }=4T^{\pm}/mc^{2}\gg1$. The parameter
$\epsilon=[G(T_{0}^{+})-G(T_{0} ^{-})]/G(T_{0}^{+})$ measures the
background temperature asymmetry of plasma species. For the
nonrelativistic temperatures
$\epsilon=5(T_{0}^{+}-T_{0}^{-})/2mc^{2}$ while in
ultrarelativistic case $\epsilon=(T_{0}^{+}-T_{0}^{-})/T_{0}^{+}$.
The numerical factor $\kappa=1/4$ for non-relativistic
temperatures ($=1/3$ for ultrarelativistic temperatures). In
deriving Eq.(\ref{B1}) with (\ref{B2}), we have assumed that the
plasma is highly transparent and the longitudinal extent of the
pulse is much shorter than its transverse dimensions. The medium
is self-focusing \ ($df/d|A|>0$) \  for $|A|$ $<1$, while it
becomes defocusing \ ($df/d|A|<0$) for higher intensities ($|A|$
$>1$).

\bigskip

To investigate the Vortex soliton solutions of Eq. (\ref{B1}), one
makes the ansatz $A=A(r)\exp(i\lambda t+im\theta)$, where the
integer \ $m$ \ defines the topological charge, and \ $\lambda$ \
is the nonlinear frequency shift. In polar coordinates, this
substitution converts Eq. (\ref{B1}) into an ordinary differential
equation. For a nonzero topological charge, the solution has a
node at the origin: \ $A_{r\rightarrow 0}\rightarrow r^{|m|}A_{0}$
where \ $A_{0}$ \ is a constant that measures the slope of $A$ at
the origin. Depending on the values of $\lambda$ and $A_{0}$ the
soliton solution is either localized
$A_{r\rightarrow\infty}\rightarrow0$ \ or delocalized
$A_{r\rightarrow\infty}\rightarrow A_{\infty}$ (where
$\lambda=f(A_{\infty})$ ). The delocalized vortex solitons are
found to exist only in the range $0<\lambda<0.25$ ($\infty
> A_{\infty}>1$) for any value of the topological charge $m$. Thus,
while near the vortex core the medium is focusing, the background
intensity of the soliton $A_{\infty}>1$ \ i.e., far beyond of the
vortex core the medium is defocusing and consequently the
background field \ is modulationally stable.

\bigskip

In what follows we consider the stability of OVS. It is well known
that single charge ($m=1$) OVSs are stable in self-defocusing
media. The multicharged vortices, however, despite their very long
life \cite{Aranson}, should eventually decay into $m=1$ vortices
\cite{Kivshar}. The last statement (which is based on general
topological consideration) should be valid for any kind of media
and will not be addressed here. For the particular system
described by Eq. (\ref{B1}), OVSs are stable against azimuthal
perturbation for any values of the allowed background field
\cite{Ber2}; the radial stability, on the other hand,  can be
established just for high intensities. Based on the results
obtained in \cite{Kim} one should expect that for the intermediate
intensities of the EM field OVSs could be unstable for our
mixed-type nonlinearity.

\bigskip

To augment our expectation it is useful to employ the hydrodynamic
analogy based on so-called Madelung transformation
$A=\rho^{1/2}\exp(i\phi)$, where $\rho=|A|^{2}$ and $\phi$ is the
real phase \cite{Rica}. With simple algebra (and changing a
time coordinate as $t\rightarrow t/2$ ), we can transform the NSE
(\ref{B1}) into the following set of two coupled equations:
\begin{equation}
\frac{\partial\rho}{\partial t}+\nabla\cdot (\,\rho\,\mathbf{V})=0
\label{B3}
\end{equation}
\begin{equation}
\frac{\partial}{\partial t}(\,\rho V_{i})+\frac{\partial}{\partial
x_{k}}(\,\rho V_{i}V_{k})=\frac{\partial\sigma_{ik}}{\partial
x_{k}} \label{B4}
\end{equation}
These equations mimic, respectively, the conservation of mass and
momentum for a compressible fluid of density $\rho$ $(=|A|^{2})$
and velocity $\mathbf{V=\nabla\phi}$ ; an effective stress tensor
is given by
\begin{equation}
\sigma_{ik}=-P\left(  \rho\right)  \delta_{ik}+\frac{\rho}{4}\frac
{\partial^{2}\left(  \ln\rho\right)  }{\partial x_{i}\partial x_{k}}
\label{B5}%
\end{equation}
where $P=-\frac{1}{2}\int \rho f^{\prime}(\rho)d\rho$ -- is the
standard hydrostatic pressure while the second term on the right
hand side of Eq.(\ref{B5}) represents the quantum stress tensor.
Thus, the Madelung fluid obeys an Euler-like equations of motion
which, due to the presence of the quantum stress term, is
drastically different from the equations describing the dynamics
of an ideal fluid. However, in certain cases when, as for
instance, for slowly varying perturbations, the quantum stress can
be ignored and even though the obtained ideal fluid Euler
equations are not entirely physically realistic, it is found to
have some important qualitative features that are helpful in
understanding the complex dynamics of the problems described by
NSE.

\bigskip

Equations (\ref{B1})-(\ref{B5}) are valid for an arbitrary
nonlinearity function $f$. For our particular case of
$f=\rho/(1+\rho)^{2}$, the equivalent hydrostatic pressure
($P(0)=0$) may be written as
\begin{equation}
P\left(  \rho\right)  =\ln\left(  1+\rho\right)
^{1/2}-\frac{\rho\left( 1+2\rho\right)  }{\left(  1+\rho\right)
^{2}} \ .
\label{B6}
\end{equation}
Notice that, for this pressure,  the sound speed \
$c_{s}=\sqrt{dP/d\rho}=\sqrt{0.5\,\rho( \rho -1)/(1+\rho )^{3}}$ \
\ "exists" only if $\rho > 1$. It should be pointed out that the
above hydrostatic pressure is positive for high densities and
becomes negative for $\rho < 2.16$. In classical hydrodynamics
\cite{Batch} negative pressures are often associated with
cavitation, which involves the formation of topological defects in
the form of bubbles. For the OVS the density $\rho=0$ at $r=0$,
and tends to its equilibrium value $\rho
_{r\rightarrow\infty}\rightarrow\rho_{0}>1$
($\rho_{0}=A_{\infty}^{2}$). Thus beyond the vortex core, the
sound can propagate, i.e. the background is modulationally stable.
The radial velocity of the fluid is zero while it performs a
differential rotation with azimuthal velocity $V_{\theta}=m/r$. If
$1<\rho _{0}<2.16$ , the hydrostatic pressure becomes negative
beyond the vortex core and the negative pressure forces may no
longer be balanced by the centrifugal and quantum pressure forces.
Thus, we expect that for the certain class of perturbation the
cavitation could take place, and since $\rho(0)$ remains zero due
to the topological constrains (strictly speaking for $m=1$) the
cavitating bubble expanding in radial direction, might be formed.

\begin{figure}
\begin{center}
\includegraphics[scale=0.37,angle=0]{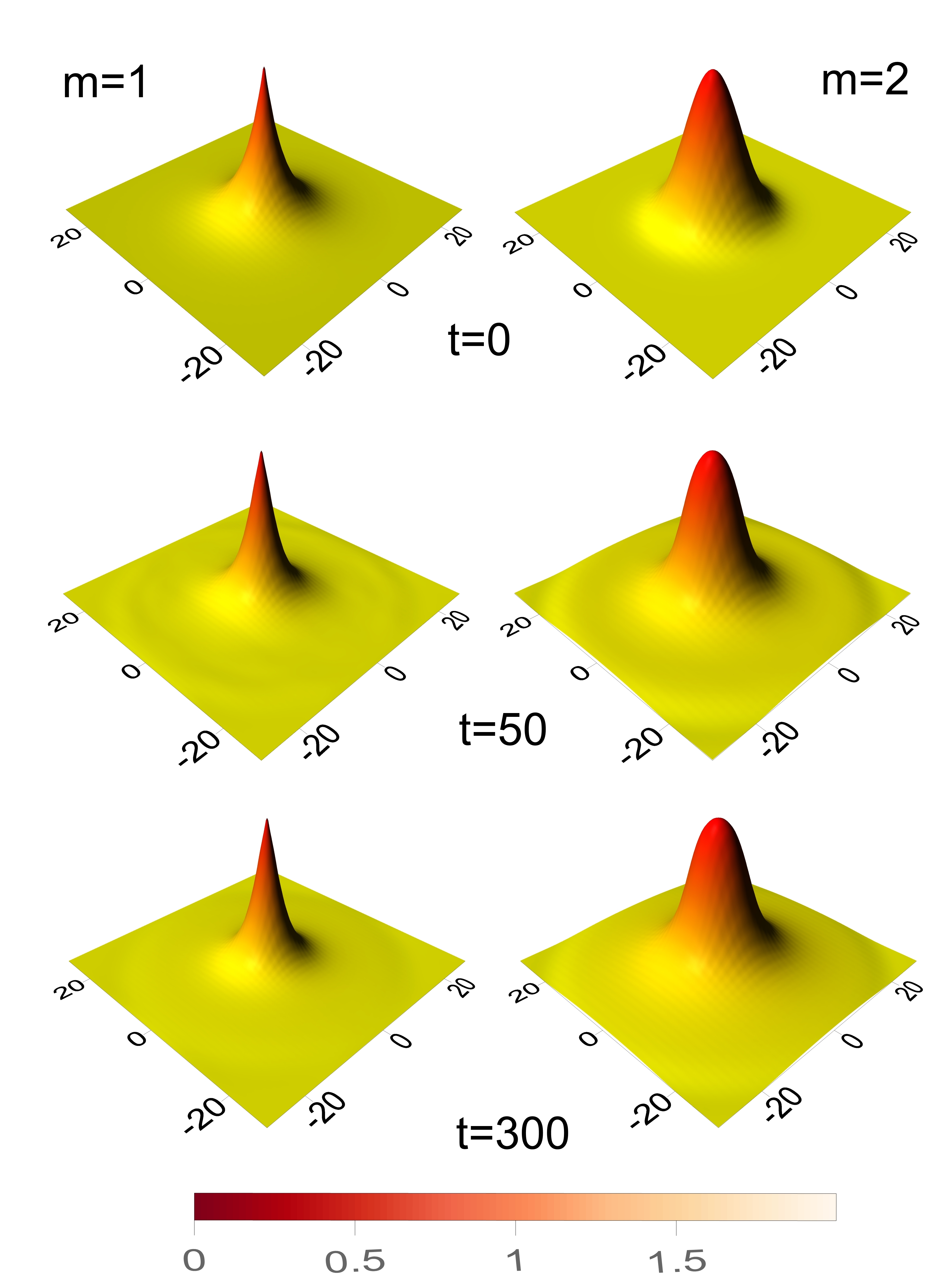}
\caption{ \ \ Dynamics of initial vortices with $A_{\infty} = 2$
($P>0$) for different $t=0$ , $t=50$  and $t=300$ time-moments.
Left column corresponds to $m=1$ and right column -- to $m=2$
cases, respectively. Spacial patterns are presented inside out for
clear view.} \label{Fig.1.}
\end{center}
\end{figure}

\bigskip

In order to verify the phenomenon suggested by the analysis of the
Madelung-fluid "translation" of the NSE, we have carried out the
stability analysis by solving Eq.(\ref{B1}) numerically. The
dynamics of initial OVS-like singly- and doubly-charged field
distribution was followed for negative $P(\rho_{0}) < 0$ ($1 <
A_{\infty}^{2}< 2.16$ ) as well as positive pressures $P(\rho_{0})
< 0$ ($A_{\infty}^{2} > 2.16$ ). For positive pressure (Fig.1),
the initial field distributions evolve toward the stationary OVS
with parameters determined by $A_{\infty}$ (plotted spacial
patterns are presented inside out for clear view). For negative
pressure (Fig.2.), on the other hand, no OVS are formed; instead,
the vortex core expands out constantly in time leading to the
generation of ever expanding electromagnetic bubbles.

\begin{figure}
\begin{center}
\includegraphics[scale=0.37,angle=0] {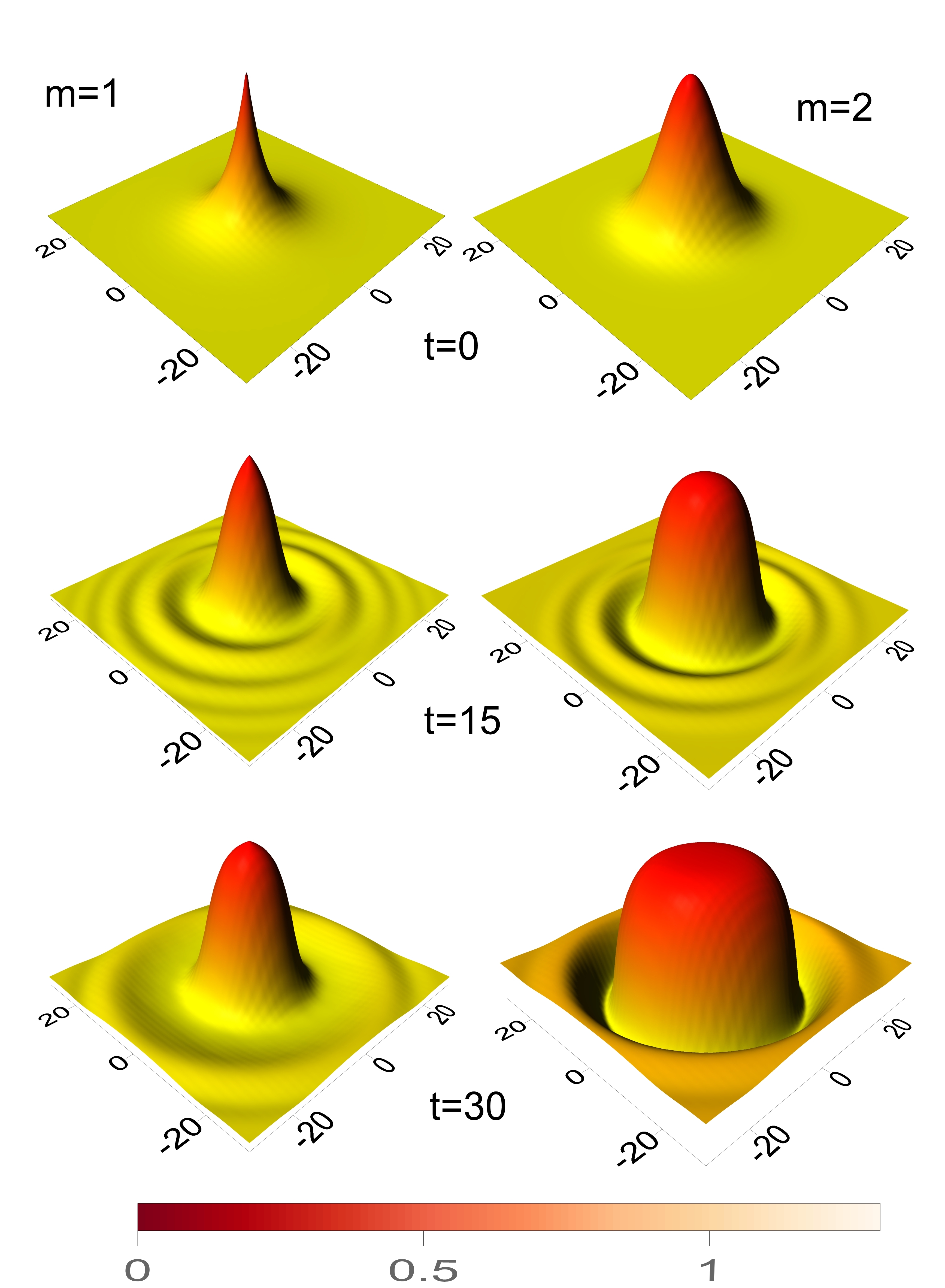}
\caption{ Dynamics of initial vortices with $A_{\infty} = 1.3$
($P<0$) for different $t=0$ , $t=15$ and $t=30$ time-moments. Left
column corresponds to $m=1$ and right column -- to $m=2$ cases,
respectively. } \label{Fig.2.}
\end{center}
\end{figure}

\bigskip

\centerline{* * *}

We  have shown that the delocalized vortex solitons associated
with an NSE, containing the particular form of saturating
nonlinearity arising naturally in a relativistic pair plasma with
temperature \textquotedblright asymmetry\textquotedblright\,, can
become unstable for moderately intense background fields leading
to the formation of an ever expanding  core. Though the changes in
the  core plasma density and temperature are small, the
electromagnetic fields are expelled  from within the expanding
core. The formation of such electromagnetic bubbles could have
rather interesting implications for structure formation both in
cosmic/astrophysical settings, and for pair plasmas that may,
soon, be available in laboratory conditions.

\bigskip

\acknowledgments{Authors acknowledge special debt to the Abdus
Salam International Centre for Theoretical Physics, Trieste,
Italy; Authors also thank Dr. Milivoj Beli\'c for valuable
discussions. The work of VIB and NLS was partially supported by
Shota Rustaveli NSF Grant project 1-4/16 (GNSF/ST09-305-4-140).
The work of SMM was supported by USDOE Contract No.~DE--FG
03-96ER-54366. The work of BNA was partially supported by the
Qatar NRF project NPRP 5-674-1-114 and by the Ministry of Science
of Serbia under the project OI 171006.}


\bigskip

\begin{thebibliography}{99}                                                                                               %

\bibitem{Kivshar} A.S. Desyatnikov and Y.S. Kivshar, Progress in Optics
\textbf{47}, 291 (2005).

\bibitem{skriabin}W.J. Firth and D. V. Skryabin, Phys. Rev. Lett. \textbf{79},
2450 (1997).

\bibitem{Ber1} V.I. Berezhiani, V. Skarka, and N.B. Aleksic, Phys. Rev. E
\textbf{64, }57601 (2001).

\bibitem{malomed} B. L. Lawrence and G. I. Stegeman, Opt. Lett.
\textbf{23}, 591 (1998); G. Boudebs, S. Cherukulappurath, H.
Leblond, J. Troles, F. Smektala, F. Sanchez, Opt. Commun.
\textbf{219}, 427 (2003); V. Skarka, N.B. Aleksi\'{c}, M. Derbazi,
and V.I. Berezhiani, Phys. Rev. B\textbf{81}, 035202 (2010).

\bibitem{PlasmaSat} P. K. Shukla, N. N. Rao, M. Y. Yu, and N. L. Tsintsadze,
Phys. Rep. \textbf{138}, 1 (1986); V. I. Berezhiani, S. M.
Mahajan, Z. Yoshida, and M. Ohhashi, Phys. Rev. E \textbf{65},
047402 (2002); V. I. Berezhiani and S. M. Mahajan, Phys. Rev.
Lett. \textbf{73}, 1110 (1994); T. Tatsuno, V. I. Berezhiani, and
S. M. Mahajan, Phys. Rev. E \textbf{63}, 046403 (2001); S. Kartal,
L. Tsintsadze and V.I. Berezhiani. Phys. Rev. E {\bf 53}, 4225
(1996).

\bibitem{Nana} S.M. Mahajan, N.L. Shatashvili, and V.I. Berezhiani, Phys. Rev.
E \textbf{80}, 066404 (2009).

\bibitem{Ber2} V.I. Berezhiani, S.M. Mahajan, N.L. Shatashvili, Phys. Rev. A
\textbf{81}, 053812 (2010); V.I. Berezhiani, S.M. Mahajan, and N.L.
Shatashvili, J. of Plasma Phys. \textbf{76}, 467 (2010).

\bibitem{Kim} W. Kim, S. Chae, and H.T. Moon, Phys. Lett. A \textbf{276}, 91 (2000).

\bibitem{Berloff} N.G. Berloff, Fluid Dyn. Res. \textbf{41}, 051403 (2009).

\bibitem{Aranson} I. Aranson and V. Steinberg, Phys. Rev. B 53, 75 (1995).

\bibitem{Rica} T. Takabayasi, Prog. Theor. Phys. \textbf{8}, 143 (1952); C.
Josserand,Y. Pomeau, and S. Rica, Phys. Rev. Lett. \textbf{75}, 3150 (1995);
\ D. Novoa, H. Michinel, and D. Tommasini, Phys.Rev. Lett. \textbf{103},
023903 (2009).

\bibitem{Batch} G. Batchelor, \textit{An Introduction to Fluid Mechanics},
Cambridge University Press, Cambridge, (1967).

\end{thebibliography}
\end{document}